\documentclass[3p,times,sort&compress]{elsarticle}

\usepackage{ecrc}
\usepackage{epstopdf}
\volume{00}
\firstpage{1}
\journalname{Nuclear Physics A}
\runauth{Andersen et al.}
\jid{nupha}
\jnltitlelogo{Nuclear Physics A}

\usepackage{graphicx}
\usepackage{amsmath,amssymb}
\usepackage[colorlinks=true,linktocpage=true,linkcolor=blue,citecolor=blue]{hyperref}

\begin{document}

\begin{frontmatter}

\title{Three loop HTL perturbation theory at finite temperature and chemical potential}

\author[a]{Michael Strickland\footnote{Presenter}$^{\!,}$}
\author[b]{Jens O. Andersen}
\author[c]{Aritra Bandyopadhyay}
\author[c]{Najmul Haque}
\author[c]{Munshi G. Mustafa}
\author[d]{Nan Su}
\address[a]{Department of Physics, Kent State University, Kent, Ohio 44242, United States}
\address[b]{Department of Physics, Norwegian University of Science and  Technology, N-7491 Trondheim, Norway}
\address[c]{Theory Division, Saha Institute of Nuclear Physics, 1/AF Bidhannagar, Kolkata-700064, India}
\address[d]{Faculty of Physics, University of Bielefeld, D-33615 Bielefeld, Germany}

\begin{abstract}
In this proceedings contribution we present a recent three-loop hard-thermal-loop perturbation theory (HTLpt) calculation of the thermodynamic potential for a finite temperature and chemical potential system of quarks and gluons. We compare the resulting pressure, trace anomaly, and diagonal/off-diagonal quark susceptibilities with lattice data. We show that there is good agreement between the three-loop HTLpt analytic result and available lattice data.
\end{abstract}


\end{frontmatter}

\section{Introduction}
\label{sect:intro}

With the advent of modern high-energy colliders the study of the quark gluon plasma (QGP) has advanced tremendously.  One outstanding question that lingers, however, is to what extent can one use ideas stemming from perturbation theory for QGP phenomenology.  One focal point in this regard has been high loop-order calculations of the equation of state of finite temperature and density QCD and comparison of these approximations to lattice QCD results.  The perturbative calculation of QGP thermodynamics has a long history~\cite{shuryak,chin,kapusta79,toimela,arnoldzhai1,arnoldzhai2,zhaikastening,braatennieto1,braatennieto2} and the perturbative expansion of the pressure of QCD at both zero~\cite{kajantie} and non-zero chemical potential~\cite{vuorinen1,vuorinen2,ipp} are now known through order $g^6\ln g$.  However, one finds in practice that a strict expansion in the coupling constant converges only for temperatures many orders of magnitude higher than those relevant for heavy-ion collision experiments.  The source of the poor convergence comes from contributions from soft momenta, $p \sim g T$.  This suggests that one needs a reorganization of finite-temperature/density perturbation theory that treats the soft sector more carefully.  

There are various ways of reorganizing the finite temperature/chemical potential perturbative series.  Here we will focus on a method called hard-thermal-loop perturbation theory (HTLpt).  For scalar field theories one can use a simpler variant called ``screened perturbation theory'' (SPT)~\cite{spt1,spt2,spt3,spt4,spt5} which was inspired in part by variational perturbation theory~\cite{vpt1,vpt2,vpt3}. A gauge-invariant generalization of SPT called HTLpt was developed by Andersen, Braaten, and Strickland over a decade ago~\cite{andersen1}.  Since then HTLpt has been used to calculate thermodynamic functions at one loop order~\cite{andersen1,andersen2,andersen3,sylvain1,sylvain2}, two loop order~\cite{andersen4,andersen5,najmul2,najmul2qns}, and three loop order at zero chemical potential~\cite{3loopglue1,3loopglue2,3loopqed,3loopqcd1,3loopqcd2,3loopqcd3} as well as at finite chemical potential(s)~\cite{najmul3,najmul4}.

\begin{figure}[t]
\centerline{
\includegraphics[width=0.44\textwidth]{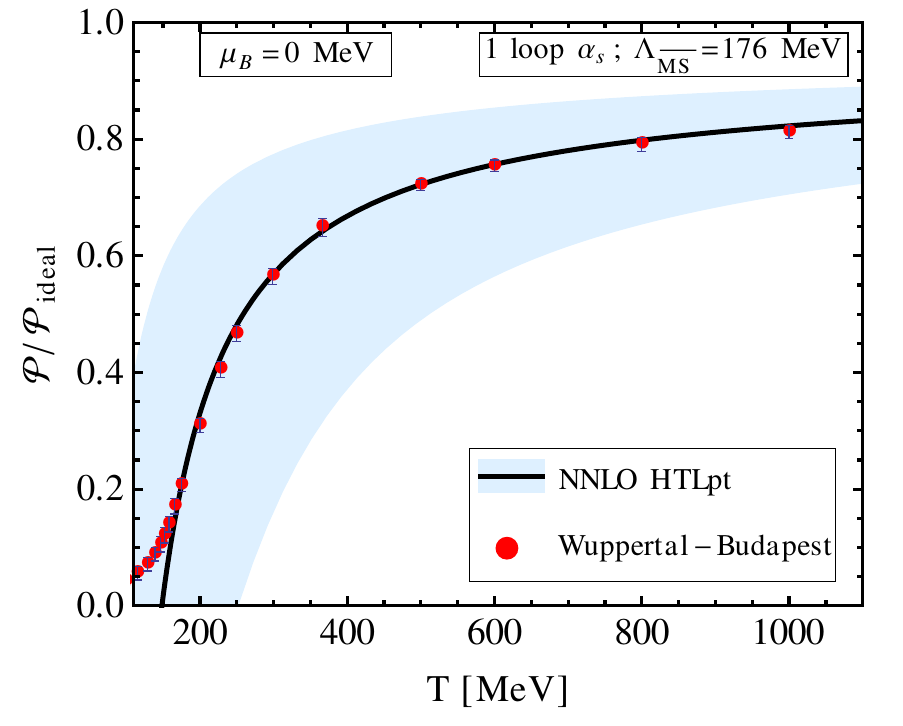}
\hspace{4mm}
\includegraphics[width=0.44\textwidth]{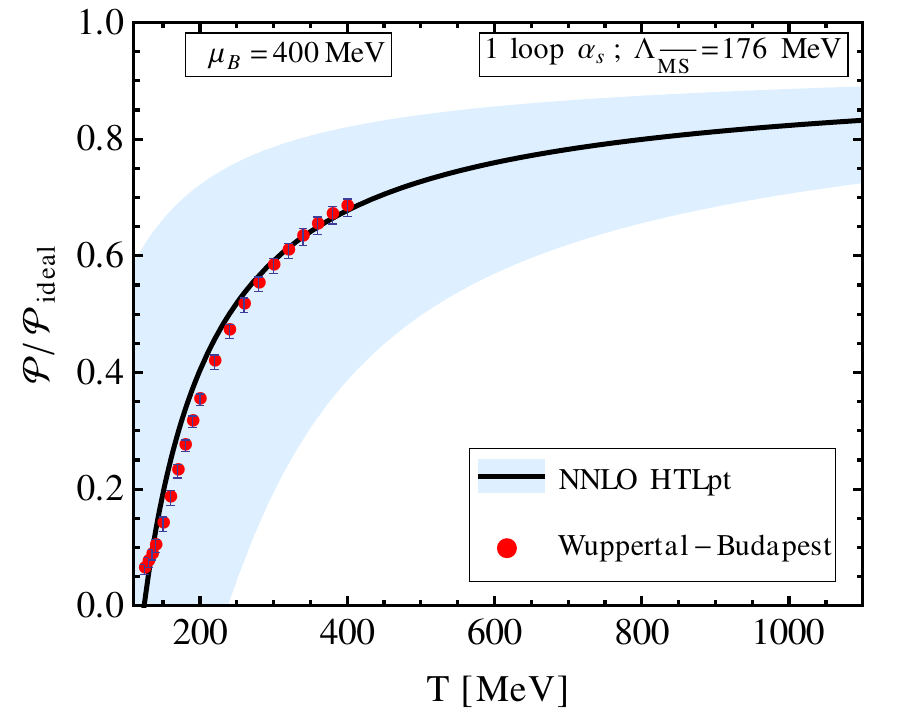}
}
\vspace{-2mm}
\caption{
Comparison of the $\mu_B=0$ (left) and $\mu_B=400$ MeV (right) NNLO HTLpt 
pressure with lattice data from Bors\'anyi et al. \cite{borsanyi1} and \cite{borsanyi3}, respectively.
}
\label{fig:pressure}
\vspace{7mm}
\centerline{
\includegraphics[width=0.44\textwidth]{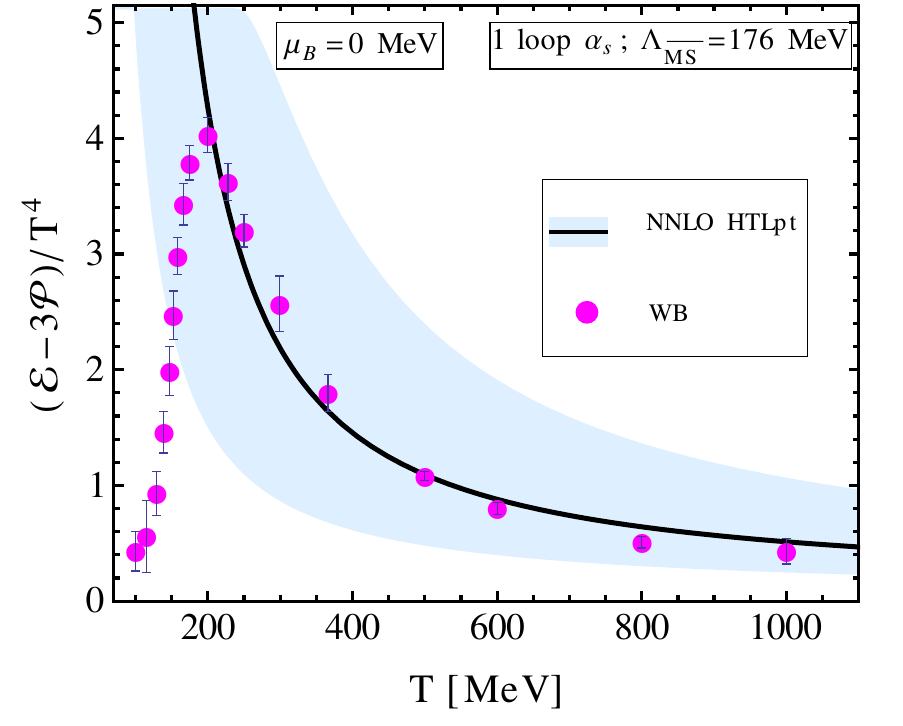}
\hspace{4mm}
\includegraphics[width=0.44\textwidth]{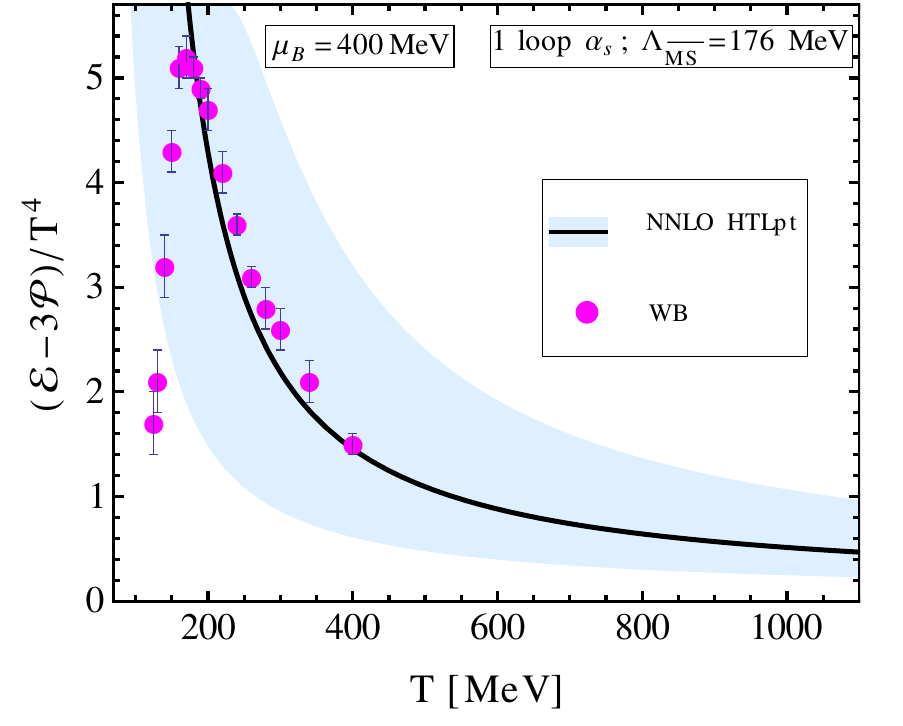}
}
\vspace{-2mm}
\caption{
Comparison of the $\mu_B=0$ (left) and $\mu_B=400$ MeV (right) NNLO HTLpt 
trace anomaly with lattice data from Bors\'anyi et al. \cite{borsanyi1} and \cite{borsanyi3}, respectively. 
}
\label{fig:ta}
\end{figure}

\begin{figure}[t]
\centerline{
\includegraphics[width=0.44\textwidth]{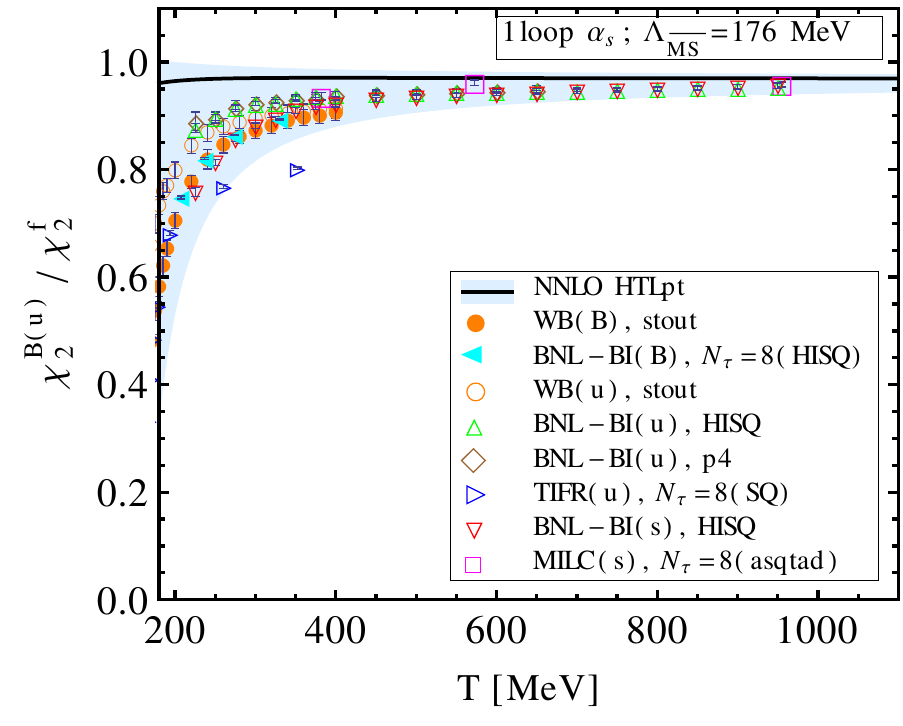}
\hspace{4mm}
\includegraphics[width=0.44\textwidth]{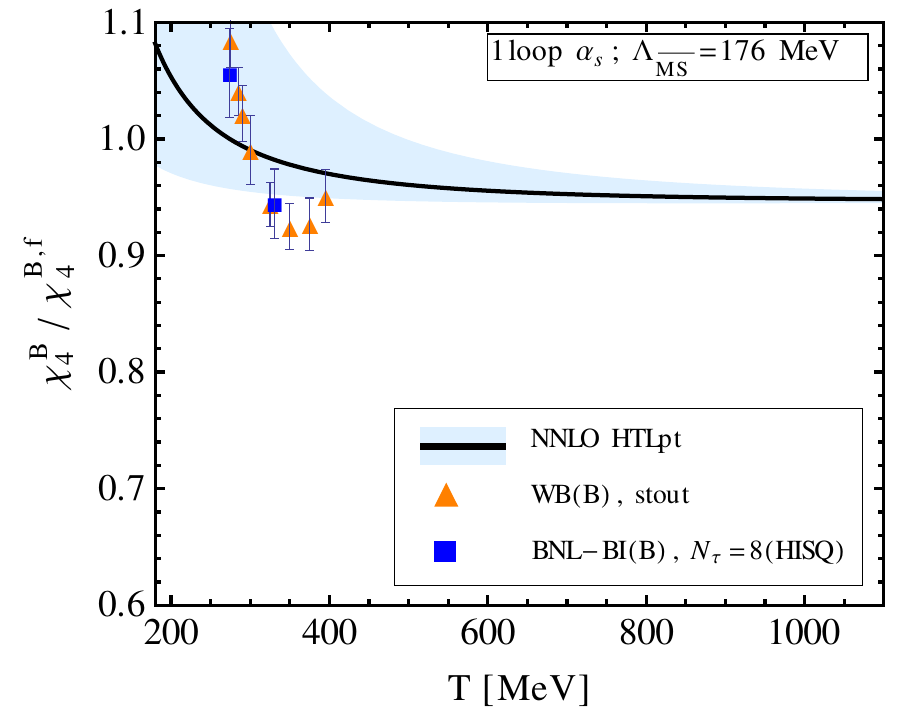}
}
\vspace{-2mm}
\caption{
The scaled second-order (left) and fourth-order (right) baryon number susceptibilities compared with various lattice data. The lattice data labeled WB, BNL-BI(B), BNL-BI(u,s), MILC, and TIFR come from Refs.~\cite{borsanyi2}, \cite{bnlb1}, \cite{bnlb2}, \cite{milc}, and \cite{TIFR}, respectively.
}
\label{fig:bns}
\vspace{7mm}
\centerline{
\includegraphics[width=0.44\textwidth]{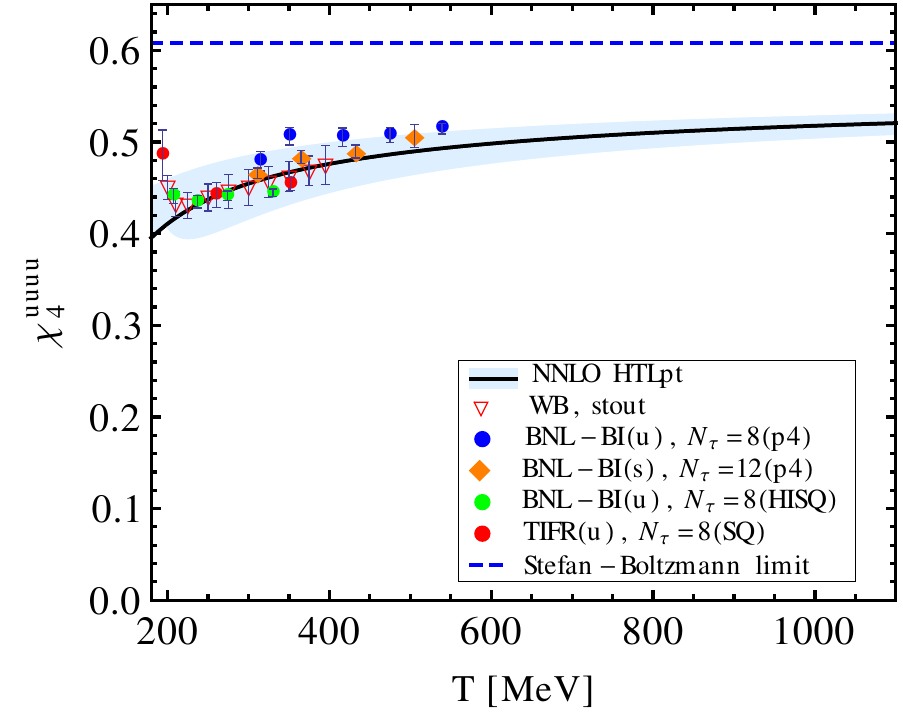}
\hspace{4mm}
\includegraphics[width=0.44\textwidth]{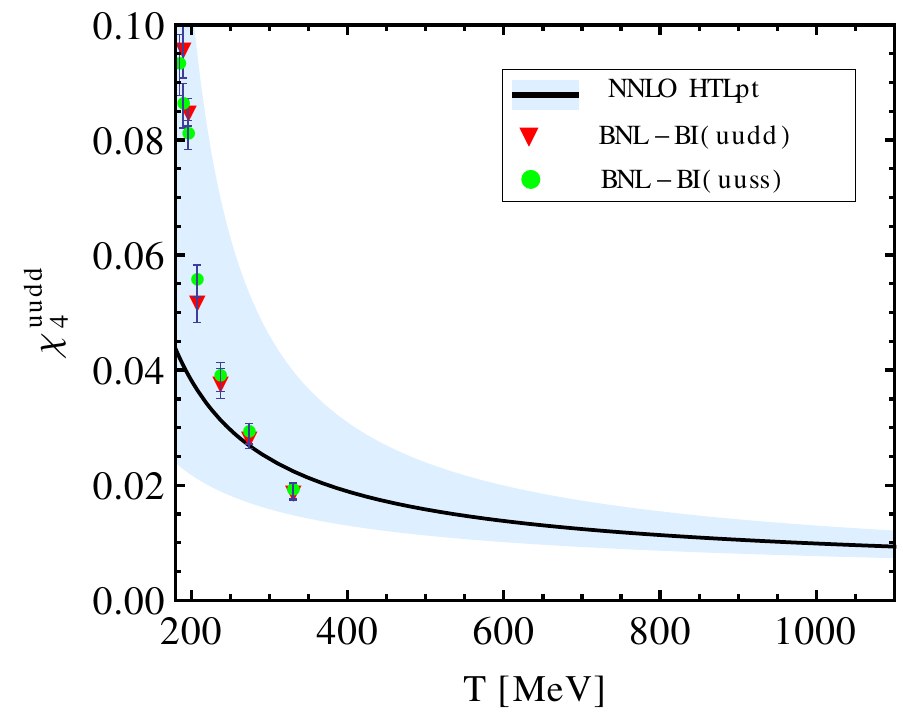}
}
\vspace{-2mm}
\caption{
Comparison of the NNLO HTLpt fourth order diagonal single quark number susceptibility (left) and the only non-vanishing fourth order off-diagonal quark number susceptibility (right) with lattice data.  In the left figure the dashed blue line indicates the Stefan-Boltzmann limit for this quantity.  The data labeled BNL-BI(uudd), BNL-BI(u,s), BNL-BI(uuss), and TIFR come from Refs.~\cite{bnlb1}, \cite{bnlb2}, \cite{bnlb3}, and \cite{TIFR}, respectively.
}
\label{fig:qns}
\end{figure}

In this proceedings contribution we present a recent calculation of the thermodynamic potential at finite temperature and chemical potential(s) to three-loop order (next-to-next-to-leading order or NNLO) in HTLpt.  The result for equal quark chemical potentials was first presented in Ref.~\cite{najmul3} and the extension to flavor-dependent chemical potentials was presented in Ref.~\cite{najmul4}.  In both cases, the resulting three-loop thermodynamic potential is renormalized using only known vacuum, mass, and coupling constant counterterms and the final result is completely analytic and gauge invariant.  The resulting analytic thermodynamic potential can be used to obtain, for example, the pressure, energy density, entropy density, trace anomaly, speed of sound, and various quark number susceptibilities.  As we will show below, there is good agreement between our NNLO HTLpt results and lattice data down to temperatures on the order of 300 MeV.  Below we present plots of some of our main results and refer the reader to Ref.~\cite{najmul4} for the calculation details and a more detailed discussion of the systematic uncertainties, etc.

\section{Results}
\label{sect:results}

In this section we present some of the final results from Ref.~\cite{najmul4}.  For all results shown we used the one-loop running coupling.  We fixed the scale $\Lambda_{\overline{\rm MS}}$ by requiring that $\alpha_s({\rm 1.5\;GeV}) = 0.326$ which is obtained from lattice measurements \cite{latticealpha}.  For one-loop running, this procedure gives $\Lambda_{\overline{\rm MS}} = 176$ MeV.  We use two separate renormalization scales, $\Lambda_g$ and $\Lambda_q$, for purely-gluonic and fermionic graphs, respectively.  We take the central values of these renormalization scales to be $\Lambda_g = 2\pi T$ and $\Lambda=\Lambda_q=2\pi \sqrt{T^2+\mu^2/\pi^2}$.  In all plots the thick black lines indicate the result obtained using these central values and the light-blue band indicates the variation of the result under variation of both of these scales by a factor of two, e.g. $\pi T \leq \Lambda _g \leq 4 \pi T$.  For all numerical results below we use $N_c=3$ and $N_f=3$.

In Fig.~\ref{fig:pressure} we compare the scaled NNLO HTLpt pressure for $\mu_B=0$ (left) and $\mu_B=400$ MeV (right) with lattice data.
In Fig.~\ref{fig:ta} we compare the scaled NNLO HTLpt trace anomaly for $\mu_B=0$ (left) and $\mu_B=400$ MeV (right) with lattice data.
In Fig.~\ref{fig:bns} we compare the scaled second-order (left) and fourth-order (right) baryon number susceptibilities with various lattice data.
In Fig.~\ref{fig:qns} we compare the scaled NNLO HTLpt fourth-order diagonal single quark number susceptibility (left) and the only non-vanishing fourth-order off-diagonal quark number susceptibility (right) with various lattice data.
As can be seen from Figs.~\ref{fig:pressure}-\ref{fig:qns}, the NNLO result has quite reasonable agreement with available lattice data.  For other quantities such as the energy density, higher order susceptibilities, etc. see Ref.~\cite{najmul4}.

\section{Conclusions}

In this proceedings contribution we presented results for the NNLO HTLpt QCD thermodynamic functions.  Although we did not list the explicit expression for the thermodynamic potential here due to limited space, the final result obtained in Ref.~\cite{najmul4} is completely analytic.  As can be seen from Figs.~\ref{fig:pressure}-\ref{fig:qns}, the NNLO result has quite reasonable agreement with available lattice data.  Our NNLO HTLpt result is gauge-invariant and, besides the choice of the renormalization scales $\Lambda_g$ and $\Lambda_q$, does not contain any free fit parameters.  In closing, we note that the application of hard thermal loops in the heavy ion phenomenology is ubiquitous and the fact that HTLpt is able to reproduce the finite temperature and chemical potential thermodynamic functions with reasonable accuracy offers some hope that application of this method to the computation of other quantities is perhaps not misguided.


\section*{Acknowledgments}
We thank S. Bors\'anyi, S. Datta, F. Karsch, S. Gupta, S. Mogliacci, P. Petreczky, and A. Vuorinen for useful discussions. N.H., A.B., and M.G.M. were supported by the Indian Department of Atomic Energy. M.S. was supported in part by DOE Grant No.~\mbox{DE-SC0004104}. N.S. was supported by the Bielefeld Young Researchers' Fund.


\begin{thebibliography}{99}

\bibitem{shuryak} 
E.~V. Shuryak, 
Sov. Phys. JETP {\bf 47}, 212 (1978).

\bibitem{chin} 
S.~A.~Chin,
Phys. Lett. B {\bf 78}, 552 (1978).

\bibitem{kapusta79} 
J.~I. Kapusta, 
Nucl. Phys. B {\bf 148}, 461 (1979).

\bibitem{toimela}
T. Toimela,
Phys. Lett. B {\bf 124}, 407 (1983).

\bibitem{arnoldzhai1}
P. Arnold, C.~X. Zhai, 
Phys. Rev. D {\bf 50}, 7603 (1994).

\bibitem{arnoldzhai2}
P. Arnold, C.~X. Zhai,
Phys. Rev. D {\bf 51}, 1906 (1995).

\bibitem{zhaikastening}
C.~X. Zhai and B.~M. Kastening, 
Phys. Rev. D {\bf 52}, 7232 (1995).

\bibitem{braatennieto1} 
E. Braaten and A. Nieto,
Phys.\ Rev.\ D {\bf 51}, 6990 (1995).

\bibitem{braatennieto2} 
E. Braaten and A. Nieto,
Phys.\ Rev.\ D {\bf 53}, 3421 (1996).

\bibitem{kajantie} 
K. Kajantie, M. Laine, K. Rummukainen and Y. Schroder, 
Phys. Rev. D {\bf 67}, 105008 (2003).

\bibitem{vuorinen1}
A. Vuorinen, 
Phys.\ Rev.\ D {\bf 67}, 074032 (2003).

\bibitem{vuorinen2}
A. Vuorinen,
Phys.\ Rev.\ D {\bf 68}, 054017 (2003).

\bibitem{ipp} A. Ipp, K. Kajantie, A. Rebhan, and A. Vuorinen, 
Phys.\ Rev.\ D {\bf 74}, 045016 (2006).

\bibitem{spt1}
F. Karsch, A. Patkos and P. Petreczky, 
Phys. Lett. B {\bf 401}, 69 (1997).

\bibitem{spt2}
S. Chiku and T. Hatsuda, 
Phys. Rev. D {\bf 58}, 076001 (1998).

\bibitem{spt3}
J.~O. Andersen, E. Braaten and M. Strickland, 
Phys. Rev. D {\bf 63}, 105008 (2001).

\bibitem{spt4}
J.~O. Andersen and L. Kyllingstad, 
Phys. Rev. D {\bf 78}, 076008 (2008).

\bibitem{spt5}
J.~O. Andersen and M. Strickland, 
Phys. Rev. D {\bf 64}, 105012 (2001) .

\bibitem{vpt1}
V.~I. Yukalov, 
Teor. Mat. Fiz. {\bf 26} 403, (1976).

\bibitem{vpt2}
P.~M. Stevenson, 
Phys. Rev. D {\bf 23}, 2916 (1981).

\bibitem{vpt3}
A. Duncan and M. Moshe, 
Phys. Lett. B {\bf 215}, 352 (1988).

\bibitem{andersen1} 
J.~O. Andersen, E. Braaten, and M. Strickland, 
Phys. Rev. Lett. {\bf 83}, 2139 (1999).

\bibitem{andersen2}
J.~O. Andersen, E. Braaten, and M. Strickland,  
Phys. Rev. D {\bf 61}, 014017 (1999).

\bibitem{andersen3}
J.~O. Andersen, E. Braaten, and M. Strickland, 
Phys. Rev. D {\bf 61}, 074016 (2000).

\bibitem{sylvain1} 
J.~O. Andersen, S. Mogliacci, N. Su and A. Vuorinen, 
Phys. Rev. D {\bf 87}, 074003 (2013);  

\bibitem{sylvain2} 
S. Mogliacci, J.~O. Andersen, M. Strickland, N. Su and A. Vuorinen, 
JHEP {\bf 1312}, 055 (2013).

\bibitem{andersen4}
J.~O. Andersen,  E. Braaten, E. Petitgirard, and M. Strickland,
Phys. Rev. D {\bf 66} (2002) 085016.

\bibitem{andersen5}
J.~O. Andersen,  E. Petitgirard, and M. Strickland,
Phys. Rev. D {\bf 70}, 045001 (2004).

\bibitem{najmul2}  
N.~Haque, M.~G.~Mustafa, and M.~Strickland, 
Phys.\ Rev.\ D {\bf 87}, 105007 (2013).

\bibitem{najmul2qns}
N.~Haque, M.~G.~Mustafa, and M.~Strickland, 
JHEP {\bf 1307}, 184 (2013).

\bibitem{3loopglue1}N. Su., J.~O. Andersen, and M. Strickland, 
Phys. Rev. Lett. {\bf 104}, 122003 (2010).
                    
\bibitem{3loopglue2} J.~O. Andersen, M. Strickland, and N. Su, 
JHEP {\bf 1008}, 113 (2010).

\bibitem{3loopqed} J.~O.~Andersen, M.~Strickland and N.~Su, 
Phys.\ Rev.\ D {\bf 80}, 085015 (2009).
          
\bibitem{3loopqcd1}J.~O. Andersen, L.E. Leganger, M. Strickland and N. Su, 
Phys.\ Lett.\ B {\bf 696}, 468 (2011).
                    
\bibitem{3loopqcd2}J.~O.~Andersen, L.~E.~Leganger, M.~Strickland and N.~Su, 
JHEP {\bf 1108}, 053 (2011).
                  
\bibitem{3loopqcd3}J.~O.~Andersen, L.~E.~Leganger, M.~Strickland and N.~Su, 
Phys.\ Rev.\ D {\bf 84}, 087703 (2011).

\bibitem{najmul3} 
N.~Haque, J.~O.~Andersen, M.~G.~Mustafa, M.~Strickland, and N.~Su,
Phys.\ Rev.\ D {\bf 89}, 061701 (2014).

\bibitem{najmul4} 
  N.~Haque, A.~Bandyopadhyay, J.~O.~Andersen, M.~G.~Mustafa, M.~Strickland and N.~Su,
  JHEP {\bf 1405}, 027 (2014).

\bibitem{latticealpha}
  A.~Bazavov, N.~Brambilla, X.~Garcia i Tormo, P.~Petreczky, J.~Soto and A.~Vairo,
  Phys.\ Rev.\ D {\bf 86}, 114031 (2012).

\bibitem{borsanyi1} S.~Bors\'anyi \emph{ et al.}, 
JHEP {\bf 1011}, 077 (2010).

\bibitem{borsanyi3} S.~Bors\'anyi \emph{ et al.},
JHEP {\bf 08}, 053 (2012).

\bibitem{borsanyi2} 
  S.~Bors\'anyi, Z.~Fodor, S.~D.~Katz, S.~Krieg, C.~Ratti and K.~Szabo,
  JHEP {\bf 1201}, 138 (2012)

\bibitem{bnlb1} 
  A.~Bazavov \emph{ et al.},
Phys.\ Rev.\ Lett.\  {\bf 111}, 082301 (2013).

\bibitem{bnlb2} 
A.~Bazavov {\it et al.},
  arXiv:1309.2317 [hep-lat].
      
\bibitem{milc} 
  C.~Bernard \emph{ et al.},
  Phys.\ Rev.\ D {\bf 71}, 034504 (2005).

\bibitem{bnlb3} 
  A.~Bazavov \emph{ et al.},
  Phys.\ Rev.\ Lett.\  {\bf 109}, 192302 (2012).

\bibitem{TIFR}
S.~Datta, R.~V.~Gavai and S.~Gupta,
  PoS LATTICE {\bf 2013}, 202 (2014).
          
\end{thebibliography}
\end{document}